\documentclass[conference]{IEEEtran}
\IEEEoverridecommandlockouts
\usepackage{lipsum}
\usepackage{cite}
\usepackage{flushend}
\usepackage{amsmath,amssymb,amsfonts}
\usepackage{graphicx}
\usepackage{textcomp}
\usepackage{xcolor}
\usepackage[utf8]{inputenc}
\usepackage[T1]{fontenc}
\usepackage{hyperref} 
\usepackage{algorithm}
\usepackage{algorithmic}
\usepackage{comment}
\def\BibTeX{{\rm B\kern-.05em{\sc i\kern-.025em b}\kern-.08em
    T\kern-.1667em\lower.7ex\hbox{E}\kern-.125emX}}

\usepackage{subcaption}
\usepackage{array}
\usepackage{caption}
\usepackage{booktabs}
\usepackage{multirow}
\hypersetup{colorlinks=true, linkcolor=blue, filecolor=magenta, urlcolor=cyan, citecolor=blue}

\makeatletter
\newcommand\fs@betterruled{%
  \def\@fs@cfont{\bfseries}\let\@fs@capt\floatl@ruled
  \def\@fs@pre{\vspace*{6pt}\hrule height.8pt depth0pt \kern2pt}%
  \def\@fs@post{\kern2pt\hrule\relax}%
  \def\@fs@mid{\kern2pt\hrule\kern2pt}%
  \let\@fs@iftopcapt\iftrue}
\floatstyle{betterruled}
\restylefloat{algorithm}
\makeatother

\renewcommand{\baselinestretch}{0.95}

\begin{document}

\title{A Novel CSI-RS Reporting Scheme for RIS Optimization in O-RAN-based NextG Networks}

 \author{
    \IEEEauthorblockN{Ali Fuat Sahin\IEEEauthorrefmark{1}\IEEEauthorrefmark{3},
    Sefa Kayraklik\IEEEauthorrefmark{1}\IEEEauthorrefmark{4}, 
    Ali Gorcin\IEEEauthorrefmark{1}\IEEEauthorrefmark{3}, 
    Ibrahim Hokelek\IEEEauthorrefmark{1}\IEEEauthorrefmark{3}, 
    Ertugrul Basar\IEEEauthorrefmark{5}\IEEEauthorrefmark{4},
    Halim Yanikomeroglu\IEEEauthorrefmark{7}}
    \IEEEauthorblockA{\IEEEauthorrefmark{1}Communications and Signal Processing Research (HISAR) Lab, TUBITAK BILGEM, Kocaeli, Turkiye}
    \IEEEauthorblockA{\IEEEauthorrefmark{3}Faculty of Electrical and Electronics Engineering, Istanbul Technical University, Istanbul, Turkiye}
    \IEEEauthorblockA{\IEEEauthorrefmark{4}Department of Electrical and Electronics Engineering, Koç University, Istanbul, Turkiye}
    \IEEEauthorblockA{\IEEEauthorrefmark{5}Department of Electrical Engineering, Tampere University, Tampere, Finland}
    \IEEEauthorblockA{\IEEEauthorrefmark{7}Non-Terrestrial Networks (NTN) Lab, Systems and Computer Engineering, Carleton University, Ottawa, ON, Canada}
    \IEEEauthorblockA{Email: \{ali.sahin, sefa.kayraklik, ali.gorcin, ibrahim.hokelek\}@tubitak.gov.tr, ertugrul.basar@tuni.fi, halim@sce.carleton.ca} 
}   

\maketitle

\begin{abstract}
Reconfigurable intelligent surface (RIS) technology is a promising enabler for next-generation (NextG) wireless systems, capable of dynamically shaping the propagation environment. Integrating RIS within the open radio access network (O-RAN) architecture enables flexible and intelligent control of wireless links. However, practical RIS-assisted operation requires efficient acquisition and reporting of channel state information (CSI) to support real-time control from the base station side. This paper proposes a CSI reference signal (CSI-RS)-based reporting scheme for downlink complex channel information (CCI) to facilitate RIS optimization in an O-RAN-compliant environment. The proposed framework establishing CCI extraction and CSI-RS reporting procedures is experimentally validated on a real-world testbed integrating an open-source O-RAN system with an RIS prototype operating in the n78 frequency band. Existing channel estimation-based RIS optimization algorithms, including Hadamard and orthogonal matching pursuit (OMP), are tailored for integration into the O-RAN architecture. Experimental results demonstrate notable improvements in received signal power for both near and far users, highlighting the effectiveness and practical viability of the proposed scheme.
\end{abstract}


\section{Introduction}
\label{sectionI}

Reconfigurable intelligent surface (RIS) technology is one of the candidate network elements for next-generation (NextG) networks, owing to its capability to dynamically reconfigure the wireless propagation environment by adjusting the phase and/or magnitude of incident signals \cite{8796365}. Leveraging the adaptability and interoperability features of the open radio access network (O-RAN) architecture, RISs can be incorporated as functional network entities to enhance the performance of wireless communication systems. In practical deployments, the RIS control mechanism, such as configuring the RIS phase-shifts and coordinating its operation within the network, is typically handled at the base station side. Since RIS phase shift optimization approaches rely heavily on accurate channel state information (CSI), efficiently estimating RIS-assisted channels and subsequently reporting this information to the RIS control mechanism remain key challenges for practical RIS-assisted end-to-end communication systems.

Channel estimation for the RIS-assisted wireless communication systems has been extensively investigated in the literature \cite{9328485,9366805,8937491,9691275,9990575,11174349}. The authors in \cite{9328485} propose a low-overhead cascaded channel estimation scheme based on a double-structured orthogonal matching pursuit (OMP) approach that exploits the sparsity of the angular channels. In \cite{9366805}, a parallel factor decomposition is utilized for RIS-aided multi-user communications through iteratively recovering from noisy samples. A framework for RIS-enhanced orthogonal frequency-division multiplexing (OFDM) systems is proposed in \cite{8937491}, where channel estimations are addressed along with the RIS phase shift optimization, with a scalable extension later presented in \cite{9691275}. To validate the sparsity-based channel estimation and RIS optimization methods, both simulation and measurement experiments are reported in \cite{9990575}, while further validation under ray tracing environments is demonstrated in \cite{11174349}.

The reporting schemes defined for fifth-generation (5G) new radio (NR) networks have been adopted by many studies to enable new implementations \cite{9952605,10200612,10571079,11037049}. In \cite{9952605}, an end-to-end channel quality indicator (CQI) estimation framework based on the CSI reference signal (CSI-RS) is developed to evaluate the performance of the CSI-RS in the NR sidelink communications. The authors in  \cite{10200612} propose a deep neural network-based CSI-RS port virtualization architecture for massive multiple-input multiple-output (MIMO) systems. With its open interfaces and flexibility, O-RAN provides a platform for the CSI-RS reporting mechanism, enabling more intelligent and programmable control of network functions. To reduce overhead and computational complexity, an alternative fronthaul function split in O-RAN for CSI-RS is proposed, utilizing a least-squares (LS) channel estimation algorithm, while remaining compliant with the existing standards \cite{10571079}. Furthermore, a testbed is introduced to integrate the RIS into an end-to-end 5G O-RAN system, demonstrating improvements in users' received signal quality by applying high-level network policies \cite{11037049}. 

\begin{figure*}[t] 
    \centering
    \includegraphics[width=0.95\textwidth]{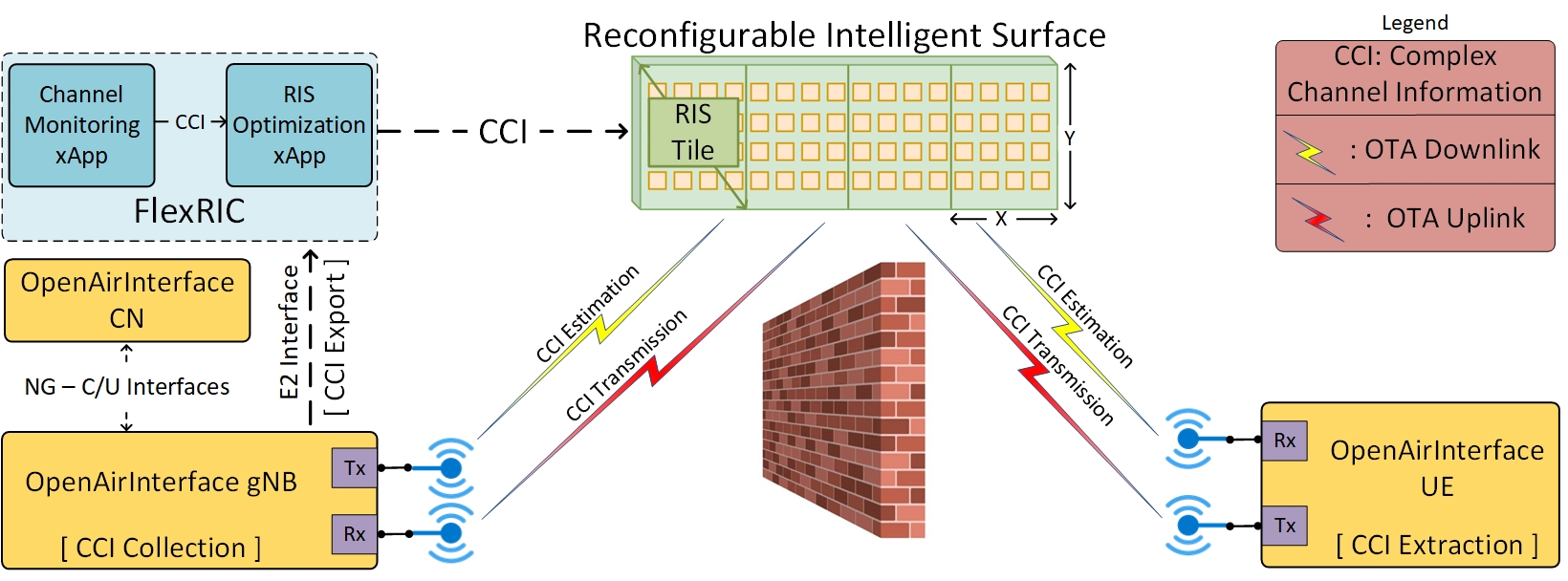}
    \caption{System model of an RIS-assisted cellular network including the CSI-RS reporting scheme for downlink CCI.}
    \label{sysMod}
    \vspace{-10 pt}
\end{figure*}

Building upon these advancements, this paper introduces a novel CSI-RS reporting scheme for downlink complex channel information (CCI) to enable practical RIS optimization within an RIS-assisted O-RAN system.
After defining the signal model of the proposed RIS-assisted O-RAN communication system, the algorithmic procedures for user CCI extraction and the associated CSI-RS reporting mechanism are presented. The system's operation is experimentally validated through over-the-air measurements on a testbed consisting of an open-source, O-RAN-compliant NextG platform integrated with an RIS prototype operating in the n78 frequency band \cite{yerliRIS}. Furthermore, channel estimation-based RIS optimization methods, including Hadamard and OMP algorithms, are adapted to practical constraints and integrated into the O-RAN architecture through xApp execution in the RIC. The experimental results demonstrate notable gains in received signal power for both near and far users, as well as the impact of CCI quantization levels on RIS optimization effectiveness.

\section{System Architecture}
\label{sectionII}

\subsection{Signal Model}
An RIS-assisted cellular network, where a NextG base station serves a single user equipment (UE) with a tile-based RIS, is illustrated in Fig. \ref{sysMod}. Each tile of the RIS consists of $X$ and $Y$ elements in vertical and horizontal directions, and the RIS is composed of $L$ tiles in total. The channel gains for a single RIS element in transmitter-RIS and RIS-receiver directions are defined as $g_{l,x,y}$ and $ d_{l,x,y}$, respectively. Here, $l$ is the RIS tile index, whereas $x$ and $y$ represent the index of the RIS cell in vertical and horizontal directions, respectively. The line-of-sight (LoS) channel is defined as $h^{\text{Tx-Rx}} \in \mathbb{C}$. Following this notation, the end-to-end channel for time instant $v$ can be given as
\begin{equation}
    h[v]= \sum_{l=1}^{L} \sum_{x=1}^{X} \sum_{y=1}^{Y} g_{l,x,y}[v] \, \Gamma_{l,x,y} \, d_{l,x,y}[v]  + h^{\text{Tx-Rx}}[v],
    \label{eq:risChannel}
\end{equation}
where $\Gamma$ represents the state of an RIS cell. If a non-line-of-sight (nLoS) setting is present with a single-tiled RIS to create an effective LoS via reflection, the channel expression can be given as
\begin{equation}
    h^{\text{Tx-RIS-Rx}}[v]= \sum_{x=1}^{X} \sum_{y=1}^{Y} \Gamma_{x,y} \, \zeta_{x,y}[v],
    \label{eq:risChannelv2}
\end{equation}
where $\zeta_{x,y}$ is the RIS channel gain for a unit cell in a single RIS tile. Following, the received signal can be written as
\begin{equation}
    r[v] =  h^{\text{Tx-RIS-Rx}}[v] x[v] + \eta[v],
    \label{eq:risReceivedSignalv2}
\end{equation}
where the transmitted signal and the complex Gaussian noise at the receiver can be denoted as $x[v]$ and $\eta[v] \sim \mathcal{N}_{\mathbb{C}} (0, \sigma_n^2)$, respectively.
Additionally, a core network (CN) is entrusted to handle mobility and authentication of the UEs. Over the base station, a near-real-time RAN intelligent controller (near-RT RIC) is embedded to develop custom applications for enhancing network capabilities. Through near-RT RIC, two control applications are operated \cite{11037049}: the first one continuously collects channel-related metrics for each UE (channel monitoring xApp, CM xApp) while the second one is responsible for utilizing the channel-related metrics for RIS optimization (RIS optimization xApp, RO xApp). 

\begin{figure}[t] 
    \centering
    \includegraphics[width=0.5\textwidth]{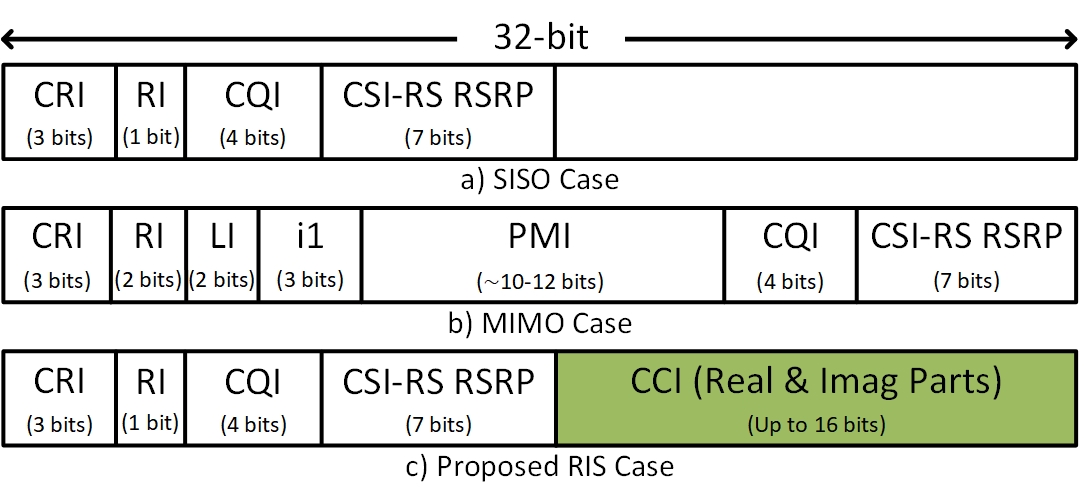}
    \caption{CSI-RS bitmaps.}
    \label{bitmap}
    \vspace{-10 pt}
\end{figure}

\subsection{System Operation}
The objective is to exploit downlink complex channel information (CCI) for RIS optimization. The system operates as follows. UE extracts CCI for the downlink channel using Algorithm \ref{alg:channel_extraction} given in Section \ref{sectionIII}. After CCI is extracted, the next step requires CCI to be transmitted through the uplink channel. To achieve this goal, a novel CSI-RS reporting configuration has been proposed. Fig. \ref{bitmap} illustrates the proposed configuration alongside the general single-input single-output (SISO) and MIMO cases. In general, the SISO case includes the CSI-RS resource indicator (CRI), rank indicator (RI), channel quality indicator (CQI), and CSI-RS reference signal received power (CSI-RS RSRP) \cite{csirsStandard}. CRI is responsible for identifying the CSI-RS resource that the UE has measured, whereas RI represents the transmission rank, and CQI indicates the link quality. Lastly, CSI-RS RSRP is the RSRP measurement for the corresponding CSI-RS grid. For the MIMO case, CRI, RI, CQI, and CSI-RS RSRP are likewise included. The only exception is that RI is constructed with 2 bits to indicate the number of transmission layers with finer granularity. Additionally, the MIMO report includes the precoding index part 1 (i1) and the precoding matrix indicator (PMI). While i1 indicates the base PMI index, PMI includes the information about the codebook type, number of ports, and such. It should be noted that the PMI length can vary with the chosen MIMO configuration. Lastly, the proposed RIS case mirrors the SISO case with the addition of CCI. While CSI-RS reports have no fixed standard limit, the practical constraint derives from uplink control information (UCI) on PUCCH \cite{uciStandard}. 3GPP approximately allows $\sim 20-50$ bits for the PUCCH Format 2, whereas the PUCCH Format 3 can allow longer UCI with approximate lengths of $\sim 160-200$ bits. To cover as many general cases as possible, we cap the CSI-RS report length at 32 bits in this study. Accordingly, the CCI payload can vary from 2 bits to 16 bits for the proposed RIS case, including the real and imaginary parts. Utilizing the proposed configuration, UE sends the CCI through the uplink channel. Next, the base station collects the CCI and then forwards it to near-RT RIC through the E2 interface. In near-RT RIC, CCI is gathered by CM xApp, which also lists other channel-related metrics (i.e., CSI-RSRP) alongside CCI. Lastly, CCI and other metrics are delivered to RO xApp to be utilized for RIS optimization algorithms given in Section \ref{sectionIII}.

\subsection{Measurement Setup}
\label{subsec:meas_setup}

The core network (CN) and NextG base station (gNB) are implemented using OpenAirInterface (OAI) \cite{oai}, which is an open-source 5G NR stack that supports both monolithic and disaggregated operation, and exposes standard 3GPP interfaces (e.g., F1-C/U, NG-C/U). Using OAI, frequency band 78 (n78, Time Division Duplexing, $f_c$ = 3.5 GHz) has been utilized for the communication. Radio access is realized using USRP B210 software-defined radios at both the gNB and the UE, configured for band n78. To ensure the blockage between the gNB and the UE, directional antennas with a half-beamwidth of $40^\circ$ are utilized in both the gNB's transmitter and the UE's receiver to steer the downlink beam. The RIS prototype with $8\times 8$ reflecting elements of binary phase shift adjustment, as a $180^\circ$ difference, is utilized at the n78 frequency bands \cite{yerliRIS}. Lastly, FlexRIC is employed for the near-RT RIC software \cite{flexric}. Developed by the OAI team, FlexRIC is an open-source, monolithic near-RT RIC software that provides an E2 interface and a programmable framework for custom xApps. The two xApps used in this work are implemented on top of FlexRIC.

\section{Methodology}
\label{sectionIII}
\subsection{Complex Channel Extraction}

\begin{figure}[t] 
    \centering
    \includegraphics[width=0.5\textwidth]{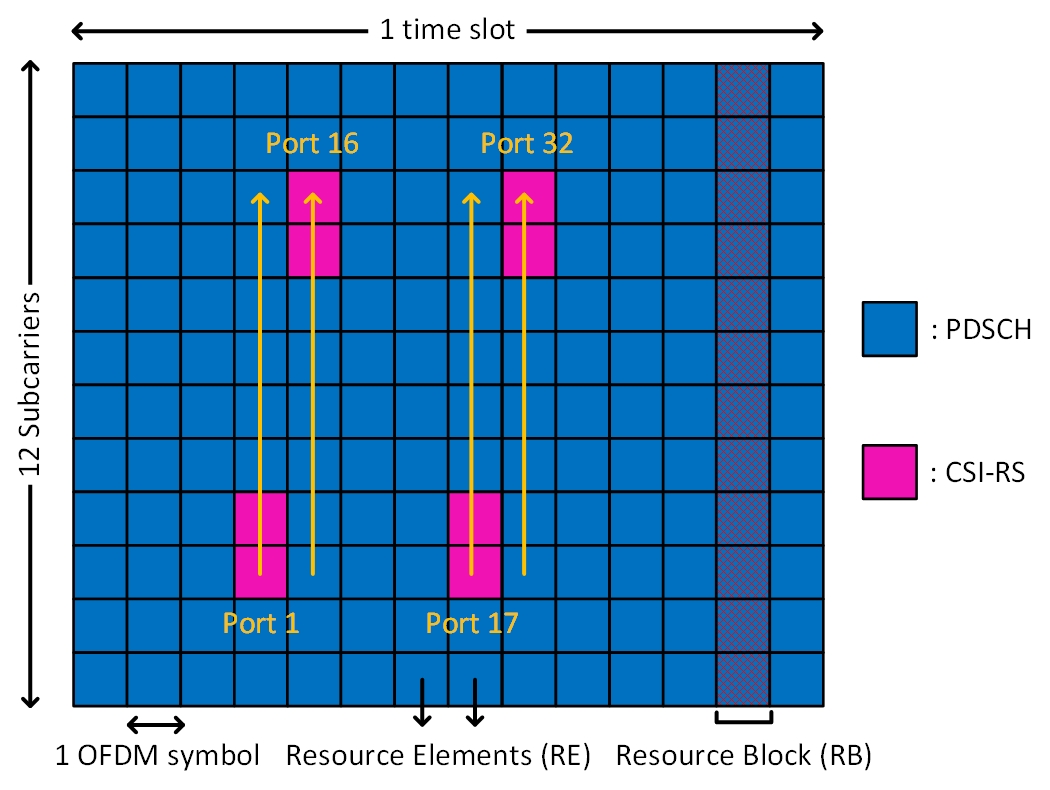}
    \caption{5G NR downlink CSI-RS grids.}
    \label{grid}
    \vspace{-10 pt}
\end{figure}

The downlink CSI-RS grid in 5G NR is depicted in Fig. \ref{grid}. CSI-RS symbols are transmitted on predefined resource elements (REs) determined by the CSI-RS configuration, code division multiplexing (CDM) configuration, and antenna ports. These pilot locations enable the UE to estimate the downlink channel for the corresponding CSI-RS resources \cite{gridStandard}. Per 3GPP mapping, it is shown that CSI-RS REs are generated with respect to row configuration and CDM groups. For each scheduled RB, the coordinates for CSI-RS RE are
\begin{equation}
    m = m_{RB} + \overline{k}_j +k' \; , \; n = \overline{l}_j + l',
    \label{eq:coordinates}
\end{equation}
where $k'$, $l'$ are the intra-pattern offsets and $\overline{k}_j$, $\overline{l}_j$ are the base frequency and time offsets for the CDM group $j$. Additionally, $m_{RB}$ is the starting subcarrier index. Hence, $m$ and $n$ can be given as the indices of the subcarrier and OFDM symbol, respectively. Moreover, let us denote the known CSI-RS symbol on the CSI-RS port $p$ as $x_p[m,n]$. Similarly, the received RE on the receiver antenna $r$ can be denoted as $Y_r[m,n]$. Following this notation, \eqref{eq:risReceivedSignalv2} can be written as
\begin{equation}
    Y_r{[m,n]} = h_r^{\text{Tx-RIS-RX}}[m,n]x_p[m,n] + \eta_{r}{[m,n]}.
    \label{eq:risReceivedSignalv3}
\end{equation}
After $Y_r{[m,n]}$ is received at the UE side, UE utilizes the CSI-RS configuration, CDM type, and bandwidth part (BWP) limits to enumerate all CSI-RS REs on port $p$ to determine the RE set $S_p$ as
\begin{equation}
    S_p = \{[m,n] \; | \; \forall \; m,n \; \text{satisfying the conditions}\}.
    \label{eq:setRE}
\end{equation}
After $S_p$ is determined, LS channel estimation is performed as
\begin{equation}
    \hat{H}^{\mathrm{LS}}_{r,p}[m,n] = \frac{Y_r{[m,n]} \, x_p^*[m,n]}{|x_p[m,n]|^2} \;\; , \;\; \forall \; [m,n] \; \epsilon \; S_p.
    \label{eq:leastSquare}
\end{equation}
With the help of LS channel estimation, the pilot can be removed, leaving the sole CCI for the given CSI-RS RE. Following the channel estimation, LS estimates are aggregated into a single anchor subcarrier $m_0$ for the given RB as
\begin{equation}
    \tilde{H}_{r,p}[m_0] = \frac{1}{N_{\text{grp}}} \displaystyle\sum_{j,s,l',k'} \hat{H}^{\mathrm{LS}}_{r,p}[m_{\mathrm{RB}} + \bar{k}_j + k',\; \bar{l}_j + l'],
    \label{eq:anchoring2}
\end{equation}
where $\sum_{j,s,l',k'} \equiv \sum_{j} \sum_{s \in \text{CDM}} \sum_{l'} \sum_{k'}$ and $N_{grp}$ is the number of LS estimates summed over. With the anchoring operation, the CCI information is averaged over an RB from a single CSI-RS RE to reduce pilot noise variance. The next step advances the averaging over the frequency domain with the interpolation operation described as
\begin{equation}
    \hat{H}_{r,p}[k] \approx \sum_{i \in \mathcal{I}} f_i[k]\, \tilde{H}_{r,p}(\text{anchor near } [k + i]),
    \label{eq:interpolation}
\end{equation}
where  $k$ is the subcarrier index over the whole BWP. Additionally, $f_i{[k]}$ represents the interpolation weights and $\mathcal{I}$ stands for a symmetric small tap set (i.e., 24 taps). Following, the wideband averaging operation is implemented such that a complex mean across the BWP is realized as
\begin{equation}
    \overline{H}_{r,p} \triangleq \frac{1}{|\mathcal{K}_{\mathrm{BWP}}|} \sum_{k \in \mathcal{K}_{\mathrm{BWP}}} \hat{H}_{r,p}[k],
    \label{eq:widebandAverage}
\end{equation}
where $\mathcal{K}_{\mathrm{BWP}}$ is the set of subcarrier indices for the corresponding BWP. The preceding processing steps follow the CSI-RS-based procedures implemented in OAI in alignment with 3GPP specifications \cite{oai, gridStandard}. Subsequently, $\overline{H}_{r,p}$ is quantized to comply with the limitations imposed by the UCI format. The quantization operation is defined as
\begin{equation}
    Q_b[\overline{H}_{r,p}] = \mathrm{clip} \left( \left\lfloor \frac{\{\overline{H}_{r,p}\}}{\Delta} \right\rceil,\, -2^{b-1},\, 2^{b-1} - 1 \right),
    \label{eq:quantization}
\end{equation}
where \textit{clip} operation enforces bounds to obey UCI limitations, $b$ indicates the desired bit length and $\Delta$ is the quantization step. Here, real and imaginary parts are quantized separately and concatenated afterward. For simplicity, a uniform symmetric quantization scheme is adopted as a baseline for evaluating CCI quantization performance. Algorithm \ref{alg:channel_extraction} summarizes the steps for complex channel extraction. After the quantization step is concluded, the proposed CSI-RS scheme given in Fig. \ref{bitmap} can be used to transmit the CCI through the uplink channel. 

\begin{algorithm}[t]
\caption{Complex Channel Extraction with CSI-RS}
\begin{algorithmic}[1]
\renewcommand{\algorithmicrequire}{\textbf{Input:}}
\renewcommand{\algorithmicensure}{\textbf{Output:}}
\REQUIRE BWP, CSI-RS config, CDM type, $Y_r[m,n]$
\ENSURE $Q_b[\overline{H}_{r,p}]$
\STATE Determine RE set $\mathcal{S}_p$ for each port $p$ using CSI-RS config from received grid $Y_r[m,n]$ \eqref{eq:setRE}.
\FOR{each $[m,n] \in \mathcal{S}_p$}
    \STATE Compute LS estimate \eqref{eq:leastSquare}.
\ENDFOR
\FOR{anchor subcarriers $m_0$ in BWP}
    \STATE Group REs to anchor $m_0$ \eqref{eq:anchoring2}. 
\ENDFOR
\STATE Interpolate across the frequency domain \eqref{eq:interpolation}.
\STATE Compute wideband average \eqref{eq:widebandAverage}.
\STATE Quantize $\Re\{\overline{H}_{r,p}\}$, $\Im\{\overline{H}_{r,p}\}$ to target bit depth \eqref{eq:quantization}.
\STATE Export quantized channel information with CSI-RS.
\RETURN $Q_b[\overline{H}_{r,p}]$
\end{algorithmic}
\label{alg:channel_extraction}
\end{algorithm}
\setlength{\textfloatsep}{10 pt}

\subsection{RIS Optimization}

\begin{algorithm}[t]
\caption{Practical Hadamard Algorithm }
\label{alg:hadamard}
\begin{algorithmic}[1]
\renewcommand{\algorithmicrequire}{\textbf{Input:}}
\renewcommand{\algorithmicensure}{\textbf{Output:}}
\REQUIRE Hadamard configuration $(\mathbf{H}_P)$, RIS config $(X,Y)$.
\ENSURE Phase-bit RIS configuration $(\boldsymbol{\beta})$.
\STATE Build sensing matrix $(\mathbf{\overline{Q}})$ with Hadamard config $\mathbf{H}_P$.
\STATE Initialize $\mathbf{y}\in\mathbb{C}^{P}$, with $P=XY$.
\FOR{$i=1$ to $P$}
  \STATE Take the $i$th row of sensing matrix $(\mathbf{Q}_{(i,:)})$ as $\mathbf{q}$.
  \STATE Map $\mathbf{q}$ to RIS pattern $(\mathbf{b} = \mathbb{1}[\mathbf{q}<0])$ and apply $(\mathbf{b})$.
  \STATE Store CSI-RS measurement $(z)$ into sensing vector $(y_i)$.
\ENDFOR
\STATE Take closed-form Hadamard inverse: $\widehat{\mathbf{g}} = \mathbf{H}^{T}\mathbf{y}/{P}$.
\STATE Remove global phase: $\phi = \angle\!\left(\sum_{i=1}^{P}\widehat{g}_i\right) \rightarrow\widehat{\mathbf{g}} = \widehat{\mathbf{g}}\,e^{-j\phi}$.  
\STATE Quantize to phase bits: $(0/\pi)$: $\boldsymbol{\beta} = \mathbb{1}\!\left[\Re\{\widehat{\mathbf{g}}\}<0\right]$.
\RETURN $\boldsymbol{\beta}$
\end{algorithmic}
\label{alg:hadamard}
\end{algorithm}

With the vectorization format in \cite{9990575} and the proposed channel estimation algorithm, the channel expression given in \eqref{eq:risChannelv2} can be written as
\begin{equation}
    h^{\text{Tx-RIS-Rx}}_{r,p} = \mathbf{\Gamma}^T \boldsymbol{\zeta}_{r,p} ,
    \label{eq:chan_est_v0}
\end{equation}
where $\mathbf{\Gamma}$ and $\boldsymbol{\zeta}_{r,p}$ are the RIS reflection coefficient vector and the RIS channel gain vector obtained by the vectorization of $\Gamma_{x,y}$ and $\zeta_{x,y}$, respectively. For simplicity, let us assume the SISO case $(r=1)$ for a nLoS setting and denote the end-to-end channel estimation as $\overline{H}_{p}$. Hence, \eqref{eq:widebandAverage} can be written as
\begin{equation}
    \overline{H}_{p} = h^{\text{Tx-RIS-Rx}}_{p} = \frac{1}{XY}\mathbf{\Gamma}^T\overline{\mathbf{F}}^*\overline{\mathbf{F}}\boldsymbol{\zeta}_p
    \label{eq:chan_est_v1},
\end{equation}
where $\overline{\mathbf{F}}$ is the 2D-DFT matrix sized $XY \times XY$ for a single-RIS tile, obeying the relation $\overline{\mathbf{F}}^*\overline{\mathbf{F}} = XY\textbf{I}$. To reveal the angular domain RIS channel vector, \eqref{eq:chan_est_v1} can be shown as
\begin{equation}
    \overline{H}_{p} =  \mathbf{\Gamma}^T\overline{\mathbf{F}}^*\boldsymbol{\xi}_p,
    \label{eq:chan_est_v2}
\end{equation}
where $\boldsymbol{\xi}_p$ is the angular domain RIS channel vector. Utilizing the far-field assumption, $\boldsymbol{\xi}_p$ can be assumed as sparse \cite{9990575}. After estimating the channel with $w$ amount of training signals through RIS sweeping, the estimated channel is
\begin{equation}
    \hat{\overline{H}}_{p,w} =  \mathbf{\Gamma}_w^T\overline{\mathbf{F}}^*\boldsymbol{\xi}_p + \hat{\eta}_{p,w},
    \label{eq:chan_est_v3}
\end{equation}
where $\hat{\eta}_{p,w}$ is the estimation noise for RIS sweep. Later, \eqref{eq:chan_est_v3} can be written in matrix form as
\begin{equation}
    \hat{\overline{\textbf{H}}}_{p} =  \overline{\mathbf{R}}\; \overline{\mathbf{F}}^*\boldsymbol{\xi}_p + \mathbf{\hat{\eta}}_{p},
    \label{eq:chan_est_v4}
\end{equation}
where $\overline{\textbf{R}}$ is the RIS reflection coefficient matrix and $\mathbf{1}$ is the vector of all ones. Here, the RIS reflection coefficient matrix can be constructed as
\begin{equation}
    \overline{\textbf{R}} = \begin{bmatrix} \overline{\mathbf{Q}} \\ - \overline{\mathbf{Q}}  \end{bmatrix},
    \label{eq:sensing_matrix}
\end{equation}
where $\overline{\mathbf{Q}}$ is the sensing matrix. Here, RIS optimization can be carried out based on the channel estimates obtained from \eqref{eq:sensing_matrix}. With the different selections of $\overline{\mathbf{Q}}$, various RIS optimization methods can be constructed. In this study, Hadamard and OMP methods have been implemented \cite{9990575,11174349}.

\subsubsection*{Hadamard Algorithm}
The Practical Hadamard algorithm utilizes the Hadamard matrix to build the sensing matrix $\mathbf{\overline{Q}}$. The Hadamard matrix is a recursive orthogonal square matrix where each element is either 1 or -1. The Hadamard matrix $(\mathbf{\Upsilon})$ can be expressed as follows
\begin{equation}
    \mathbf{\Upsilon}_{1} = \begin{bmatrix} 1 & 1 \\ 1 & -1 \end{bmatrix}, \; \; \mathbf{\Upsilon}_{2i} = \begin{bmatrix} \mathbf{\Upsilon}_{2(i-1)} & \mathbf{\Upsilon}_{2(i-1)} \\\mathbf{\Upsilon}_{2(i-1)} & -\mathbf{\Upsilon}_{2(i-1)} \end{bmatrix},
    \label{eq:hadamard_matrix}
\end{equation}
where $i$ is the size index. Utilizing this expression, the sensing matrix $\mathbf{\overline{Q}}$ can be defined as
\begin{equation}
    \mathbf{\overline{Q}} = \mathbf{\Upsilon}_{P},
    \label{eq:sensing_hadamard_matrix}
\end{equation}
where $P$ is the total number of RIS elements. Algorithm \ref{alg:hadamard} outlines the steps for the Practical Hadamard algorithm.

\begin{algorithm}[t]
\caption{Practical OMP Algorithm }
\label{alg:ompv2}
\begin{algorithmic}[1]
\renewcommand{\algorithmicrequire}{\textbf{Input:}}
\renewcommand{\algorithmicensure}{\textbf{Output:}}
\REQUIRE RIS size $(X,Y)$, OMP pilots $(W)$, sparsity $(S)$.
\ENSURE Phase-bit RIS configuration $(\boldsymbol{\beta})$.

\STATE Form DFT matrix $(\mathbf{F})$ with size $P=XY$.
\STATE Build sensing matrix $(\mathbf{\overline{Q}})$ with Bernoulli distribution.
\STATE Form dictionary as $\mathbf{Z} = \mathbf{\overline{Q}}\mathbf{\overline{F}}^{*}$.
\STATE Initialize measurement vector as $\mathbf{y}\in\mathbb{C}^{W}$.
\FOR{$w=1$ to $W$} \label{line:pilotloop}
  \STATE Take the $i$th row of sensing matrix $(\mathbf{Q}_{(i,:)})$ as $\mathbf{q}$.
  \STATE Map $\mathbf{q}$ to RIS pattern $(\mathbf{b} = \mathbb{1}[\mathbf{q}<0])$ and apply $(\mathbf{b})$.
  \STATE Store CSI-RS measurement $(z)$ into sensing vector $(y_i)$.
\ENDFOR
\STATE Initialize residual ($\mathbf{r}$) as $\mathbf{r} =\mathbf{y}$ and support $(\gamma)$ as $\gamma = \emptyset$.
\FOR{$s = 1$ to $S$}
  \STATE Compute $\rho_j = {\left| z_j^{H}\mathbf{r}\right|^{2}}/{z_j^{H}z_j}$ ; $\forall$ $z_i$ (column) $\in$ $\mathbf{Z}$.
  \STATE Select atom: $j^\star = \arg\max_j \rho_j$.
  \STATE Update support: $\gamma = \gamma \cup \{j^\star\}$.
  \STATE Update active matrix: $\mathbf{Z}_\gamma = \mathbf{Z}(:,\gamma)$.
  \STATE LS estimation: $\mathbf{g}_{\mathrm{act}} = \arg\min_{\mathbf{u}} \|\mathbf{y}-\mathbf{Z}_\gamma \mathbf{u}\|_2^2$.
  \STATE Update residuals: $\mathbf{r} = \mathbf{y} - \mathbf{Z}_\gamma \mathbf{g}_{\mathrm{act}}$.
\ENDFOR
\STATE Build angular-domain channel: $\hat{\mathbf g}[\gamma] \leftarrow \mathbf g_{\text{act}}$.
\STATE Generate element-domain response: $\hat{\mathbf h} = \mathbf F^{H}\hat{\mathbf g}$.
\STATE Remove global phase: $\phi = \angle\!\left(\sum_{i=1}^{P}\widehat{h}_i\right) \rightarrow\widehat{\mathbf{h}} = \widehat{\mathbf{h}}\,e^{-j\phi}$.  
\STATE Quantize to phase bits $(0/\pi)$: $\boldsymbol{\beta} = \mathbb{1}\!\left[\Re\{\widehat{\mathbf{g}}\}<0\right]$.
\RETURN $\boldsymbol{\beta}$
\end{algorithmic}
\end{algorithm}

\subsubsection*{Orthogonal Matching Pursuit}
Orthogonal Matching Pursuit (OMP) greedily exploits channel sparsity in the angular domain to optimize RIS with lesser overhead. Algorithm \ref{alg:ompv2} summarizes the steps for the Practical OMP algorithm. The sensing matrix $\mathbf{\overline{Q}}$ is built with a square Bernoulli matrix $(\mathbf{\Phi})$ with size $P$ as follows
\begin{equation}
    \mathbf{\overline{Q}} = \mathbf{\Phi}_{P} = \begin{bmatrix} \pm 1 & \cdots \\ \cdots & \pm 1  \end{bmatrix}_{\sqrt{P} \times \sqrt{P}} \; ,
    \label{eq:bernoulli_matrix}
\end{equation}
where each element is $\pm 1$ with equal probability according to the Bernoulli distribution.  

\section{Experimental Results}
\label{sectionIV}

\subsection{Complex Channel Extraction Results}
\begin{figure}[t]
    \centering
    \begin{subfigure}{\columnwidth}
        \centering
        \includegraphics[width=\columnwidth]{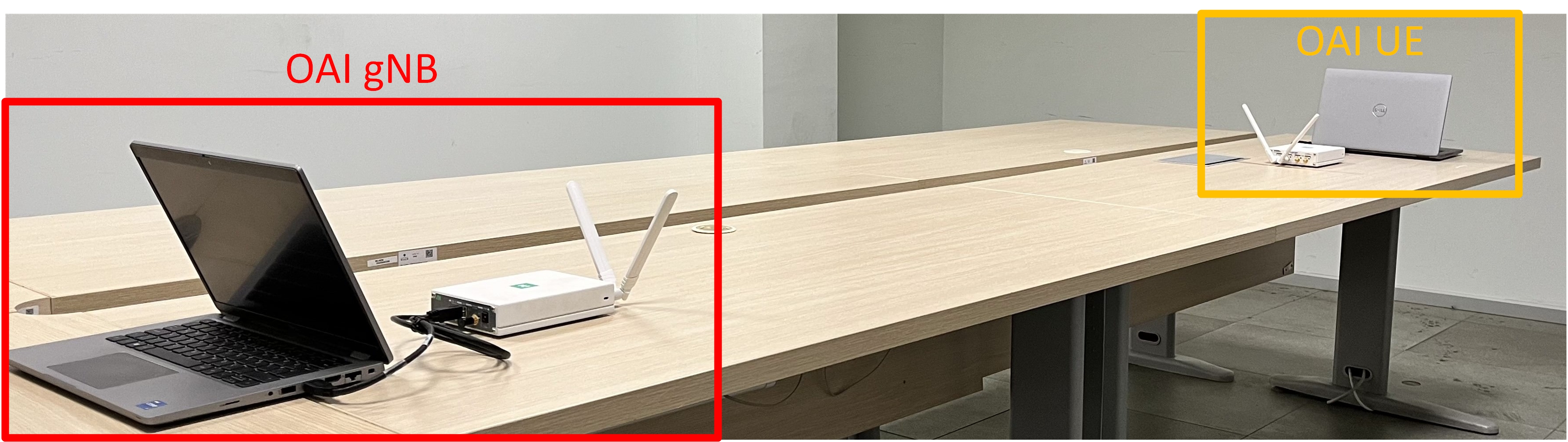}
        \caption{Measurement setup}
        \label{subfig:cci_meas_setup}
    \end{subfigure}
    \begin{subfigure}{\columnwidth}
        \centering
        \includegraphics[width=\columnwidth]{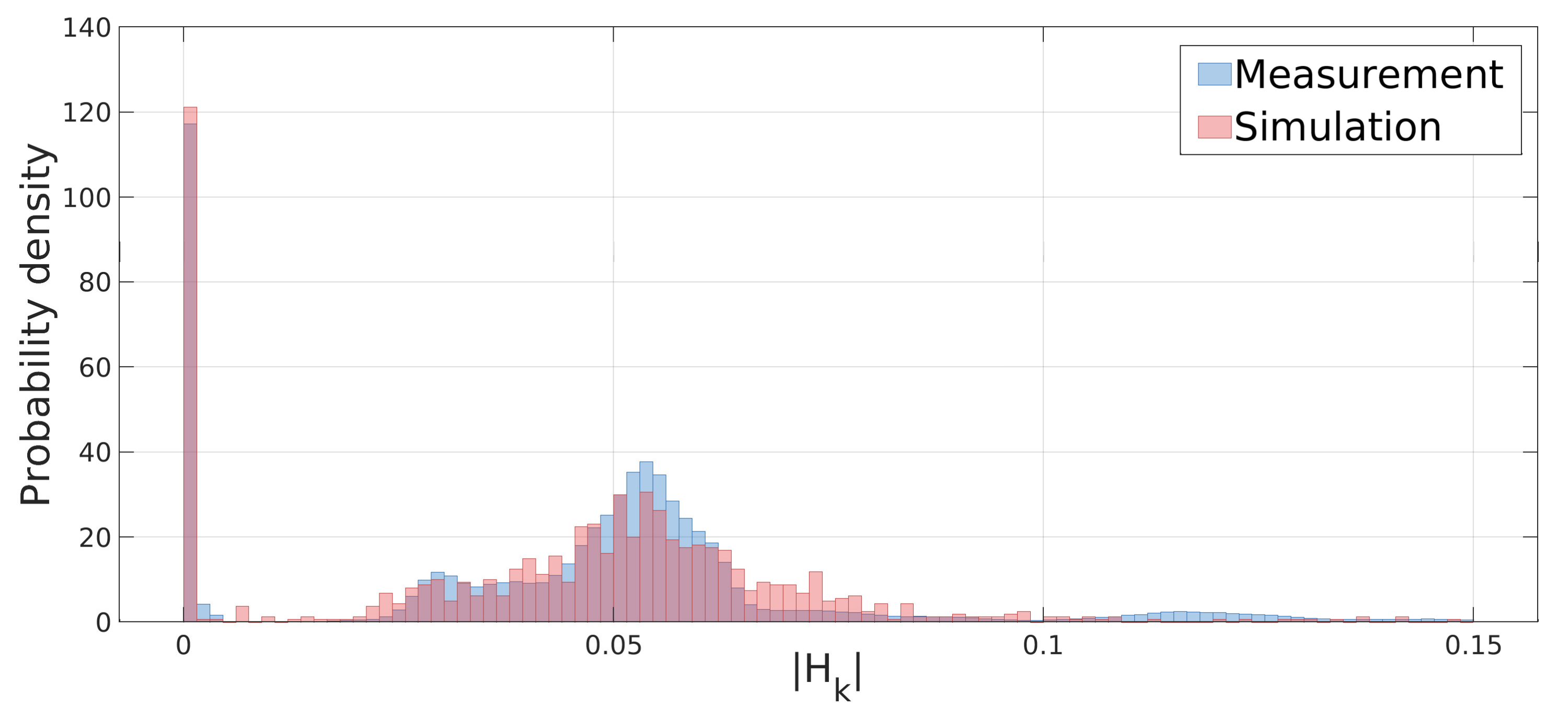}
        \vspace{-20 pt}
        \caption{Empirical PDF}
        \label{subfig:cci_meas_results}
    \end{subfigure}
    \caption{CCI validation results.}
    \vspace{-5 pt}
\end{figure}

To validate the proposed CSI-RS-based CCI reporting mechanism in a controlled single-UE setting, a simple measurement setup has been deployed, given in Fig. \ref{subfig:cci_meas_setup}. The setup employs an OAI gNB serving a single OAI UE under the LoS condition to emulate Rician fading. A 20 MHz bandwidth (106 PRBs with 30 kHz subcarrier spacing) has been allocated to the UE, and 10 Mbps downlink traffic has been generated. Fig. \ref{subfig:cci_meas_results} plots the empirical PDF of the UE-side channel-amplitude estimates collected over time. The dominant mass near $|H_k|\approx 0$ corresponds to REs that do not carry user data (guard subcarriers, unscheduled REs, and control RBs), whereas the smaller lobe around $\sim 0.06$ (after normalization) reflects active data-bearing REs. The simulation reproduces both behaviors, supporting the validity of the CCI extraction.

\subsection{RIS Optimization Results}

\begin{figure*}[t]
    \centering
    \begin{subfigure}{0.66\columnwidth}
        \centering
        \includegraphics[width=\columnwidth]{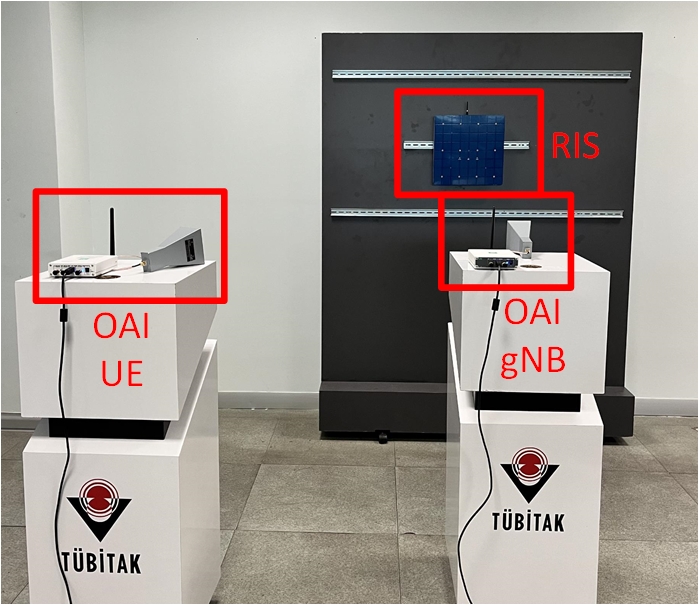}
        \caption{Measurement setup}
        \label{subfig:risOpt_setup}
    \end{subfigure}
    \hfill
    \begin{subfigure}{0.66\columnwidth}
        \centering
        \includegraphics[width=\columnwidth]{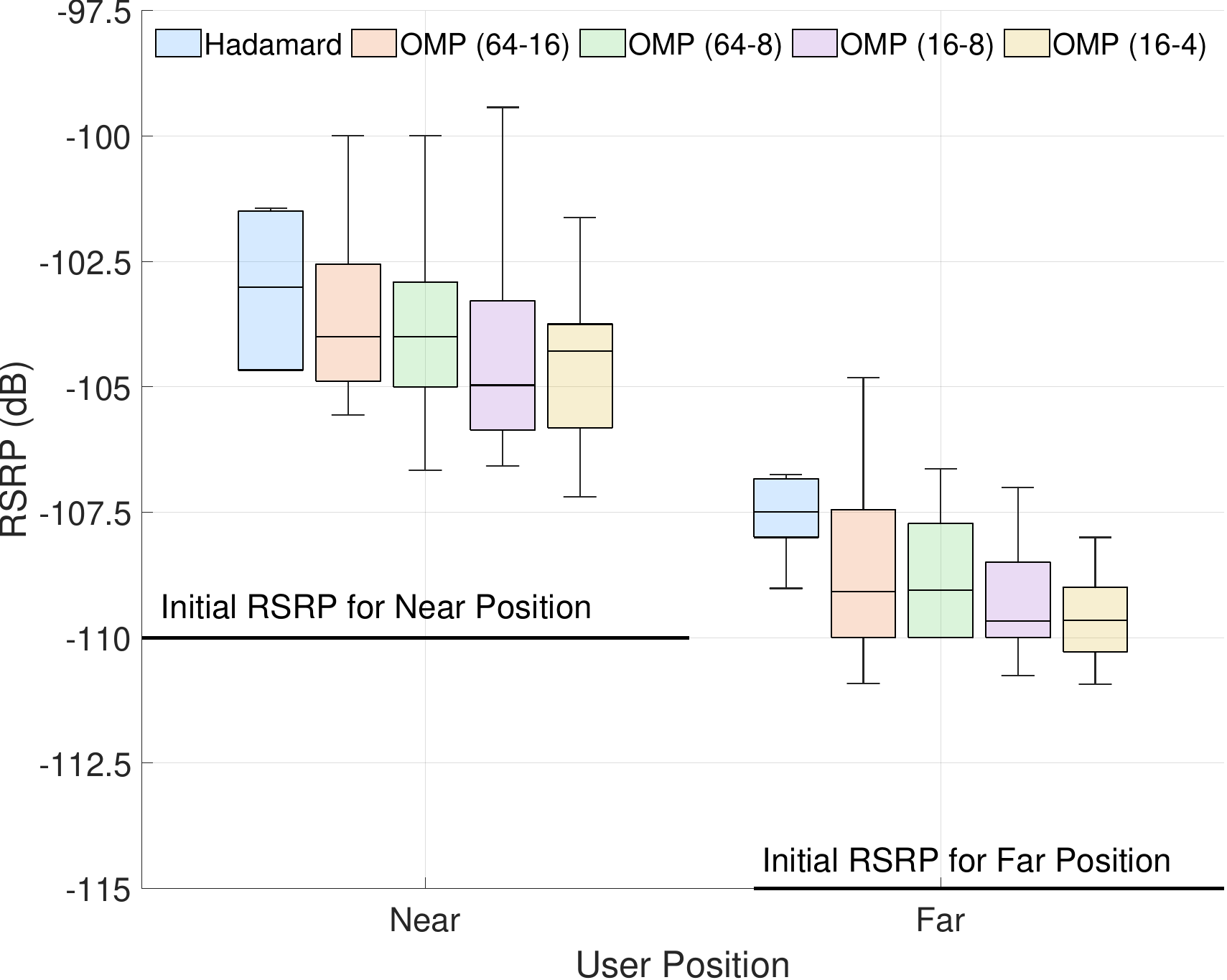}
        \caption{Location-based analysis}
        \label{subfig:risOpt_perfImp}
    \end{subfigure}
    \hfill
    \begin{subfigure}{0.66\columnwidth}
        \centering
        \includegraphics[width=\columnwidth]{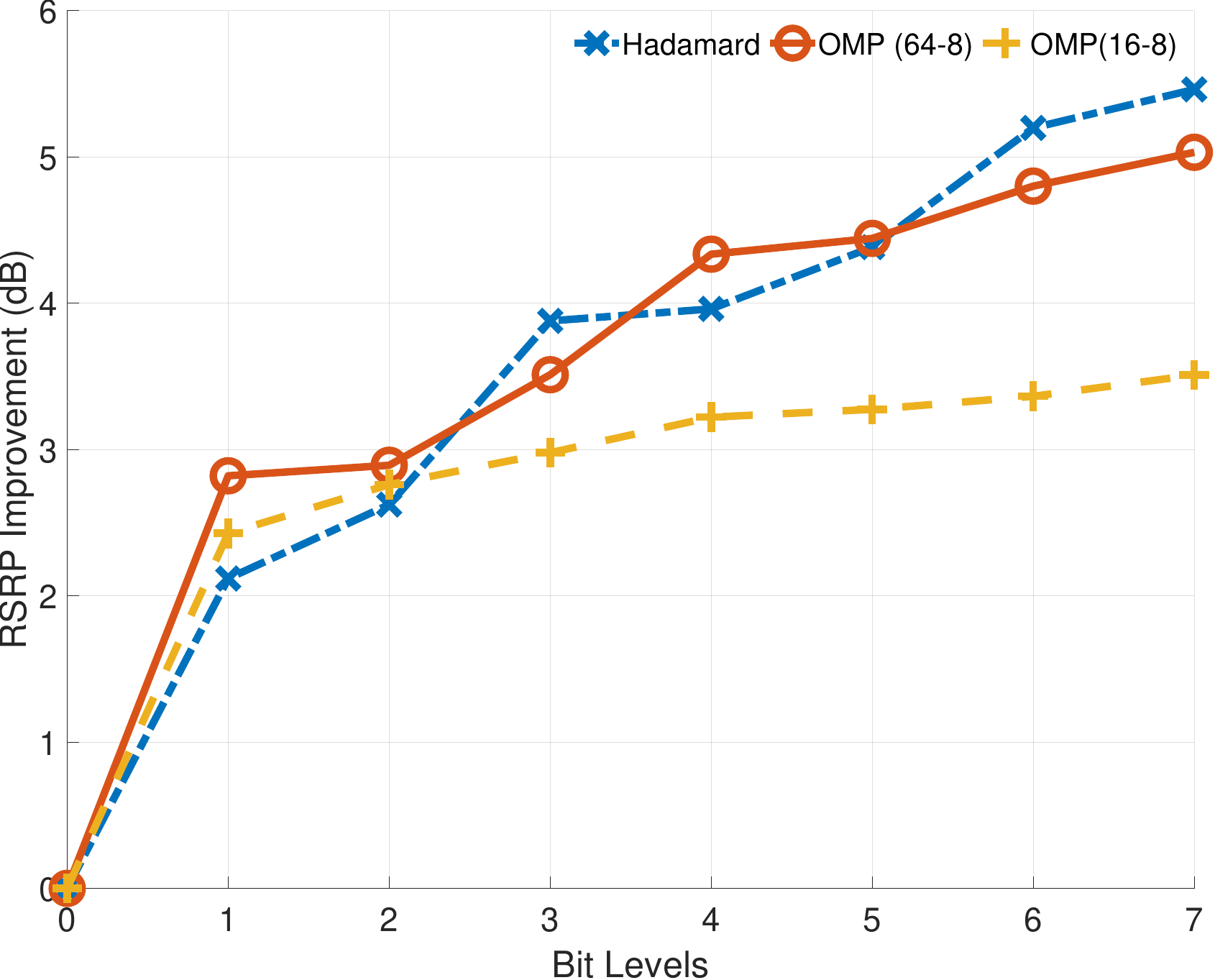}
        \caption{CCI bit length-based analysis}
        \label{subfig:risOpt_perfRes}
    \end{subfigure}
    \caption{RIS optimization measurements utilizing CCI through the CSI-RS-based reporting.}
    \label{fig:measResults}
    \vspace{-10 pt}
\end{figure*}

The practical RIS optimization setup of Section \ref{subsec:meas_setup} is shown in Fig. \ref{subfig:risOpt_setup}. We report two experiment sets using CSI-RS RSRP as the optimization metric \cite{11037049}. The first set investigated the RIS optimization performance across locations, methods, and parameters with a fixed CCI quantization length of 16 bits. The OAI UE is placed at two positions: near ($1.5$ m from the RIS) and far ($2.5$ m). The initial RSRP value is $-110$ dB (for near location) and $-115$ dB (for far location) which serve as the baseline case without RIS optimization. For OMP, we collect four configurations denoted OMP $(W,S)$, where $W$ is the number of pilots and $S$ the sparsity level. Two runs use $W=64$ with $S \in $\{8,16\}, and two use $W=16$ with $S \in $\{4,8\}. Here, each location–method pair is measured 100 times, yielding a total of 1,000 measurements. For consistency, average values are reported. The performance results are shown in Fig. \ref{subfig:risOpt_perfImp}, while the corresponding computational runtime of the considered methods is summarized in Table \ref{tab:runtimeComp}. Hadamard attains the highest improvement at both locations, closely followed by OMP(64,16) and OMP(64,8). Conversely, OMP(16,8) and OMP(16,4) exhibit significantly lower runtime with only a minor drop in gain, and the same trends hold at the far position. Lastly, the comparable performance achieved by OMP and Hadamard supports the assumption that the effective RIS-assisted channel exhibits a sparse structure. 

Fig. \ref{subfig:risOpt_perfRes} shows the second set of results where the RIS optimization performance is investigated according to CCI bit length using three methods: Hadamard, OMP(64,8), and OMP(16,8). All methods remain functional even at very low precision (e.g., 1-bit) and exhibit increasing gains as bit length increases. Hadamard and OMP(64,8) methods, which incur higher training overhead, continue to exhibit performance gains at longer bit lengths, whereas the lightweight OMP(16,8) method shows comparatively smaller performance improvements beyond 2 bits. The reported results represent the average over 25 repetitions for each method–bit-length configuration, yielding a total of 525 measurements.

\begin{table}[t]
\centering
\caption{Computational Performance Comparison}
\vspace{-5pt}
\label{tab:runtimeComp}
\begin{tabular}{
    |>{\centering\arraybackslash}m{0.17\columnwidth}|
    >{\centering\arraybackslash}m{0.105\columnwidth} 
    >{\centering\arraybackslash}m{0.105\columnwidth} 
    >{\centering\arraybackslash}m{0.105\columnwidth}  
    >{\centering\arraybackslash}m{0.105\columnwidth} 
    >{\centering\arraybackslash}m{0.105\columnwidth}|
}\hline
\textbf{Method} & Hadamard & OMP (64-16) & OMP (64-8) & OMP (16-8) & OMP (16-4)\\
\hline
\textbf{Runtime [s]} & 49.70 &  49.69 & 49.67 & 11.30 & 11.29 \\
\hline
\end{tabular}
\vspace{-5 pt}
\end{table}

\section{Conclusion \& Future Works}
\label{sectionV}
This paper presents a CSI-RS-based reporting scheme for downlink CCI, enabling efficient RIS optimization under real-time network control. The proposed architecture incorporates an open-source O-RAN platform integrated with an RIS prototype operating at the n78 frequency band. Channel estimation-based RIS optimization algorithms, including Hadamard and OMP, are adapted with practical limitations taking into consideration for the deployment within the O-RAN environment. The experiments demonstrated notable improvements in the RSRP values across various scenarios, verifying the feasibility and effectiveness of the proposed scheme. The results confirm that the combination of O-RAN’s open and programmable architecture with the RIS technology offers a promising pathway toward adaptive and intelligent NextG wireless networks. While the presented results demonstrate the feasibility of the proposed CSI-RS-based reporting framework, the current study is limited to a single-UE SISO experimental setup. Future work will investigate multi-user and MIMO scenarios in RIS-assisted O-RAN systems, including interference management, user scheduling, and resource allocation aspects, as well as advanced CCI quantization strategies and machine learning-based CSI reporting techniques.

\section*{Acknowledgment}
This paper received funding from the MOSAIC project. MOSAIC has been accepted for funding within the CHIPS Joint Undertaking, a public-private partnership in collaboration with the HORIZON Framework Programme and the national Authorities under grant agreement number 101194414. 

\bibliographystyle{IEEEtran}
\bibliography{refer}

\end{document}